\documentstyle{mn}

\newif\ifAMStwofonts

\newcommand{\gtilde}
{~ \raisebox{-1ex}{$\stackrel{\textstyle >}{\sim}$} ~}
\newcommand{\ltilde}
{~ \raisebox{-1ex}{$\stackrel{\textstyle <}{\sim}$} ~}

\ifoldfss
  \ifCUPmtlplainloaded \else
    \NewTextAlphabet{textbfit} {cmbxti10} {}
    \NewTextAlphabet{textbfss} {cmssbx10} {}
    \NewMathAlphabet{mathbfit} {cmbxti10} {} 
    \NewMathAlphabet{mathbfss} {cmssbx10} {} 
  \fi
  \ifAMStwofonts
    \ifCUPmtlplainloaded \else
      \NewSymbolFont{upmath} {eurm10}
      \NewSymbolFont{AMSa} {msam10}
      \NewMathSymbol{\upi}     {0}{upmath}{19}
      \NewMathSymbol{\umu}     {0}{upmath}{16}
      \NewMathSymbol{\upartial}{0}{upmath}{40}
      \NewMathSymbol{\leqslant}{3}{AMSa}{36}
      \NewMathSymbol{\geqslant}{3}{AMSa}{3E}

    \fi
  \fi
\fi 

\ifnfssone
  \newmathalphabet{\mathit}
  \addtoversion{normal}{\mathit}{cmr}{m}{it}
  \addtoversion{bold}{\mathit}{cmr}{bx}{it}
  \newmathalphabet{\mathbfit} 
  \addtoversion{normal}{\mathbfit}{cmr}{bx}{it}
  \addtoversion{bold}{\mathbfit}{cmr}{bx}{it}
  \newmathalphabet{\mathbfss} 
  \addtoversion{normal}{\mathbfss}{cmss}{bx}{n}
  \addtoversion{bold}{\mathbfss}{cmss}{bx}{n}
  \ifAMStwofonts
    \ifCUPmtlplainloaded \else
      \UseAMStwoboldmath
      \makeatletter
      \new@mathgroup\upmath@group
      \define@mathgroup\mv@normal\upmath@group{eur}{m}{n}
      \define@mathgroup\mv@bold\upmath@group{eur}{b}{n}
      \edef\UPM{\hexnumber\upmath@group}
      \new@mathgroup\amsa@group
      \define@mathgroup\mv@normal\amsa@group{msa}{m}{n}
      \define@mathgroup\mv@bold\amsa@group{msa}{m}{n}
      \edef\AMSa{\hexnumber\amsa@group}
      \makeatother
      \mathchardef\upi="0\UPM19
      \mathchardef\umu="0\UPM16
      \mathchardef\upartial="0\UPM40
      \mathchardef\leqslant="3\AMSa36
      \mathchardef\geqslant="3\AMSa3E
    \fi
  \fi
\fi 

\ifnfsstwo
  \DeclareMathAlphabet{\mathbfit}{OT1}{cmr}{bx}{it}
  \SetMathAlphabet\mathbfit{bold}{OT1}{cmr}{bx}{it}
  \DeclareMathAlphabet{\mathbfss}{OT1}{cmss}{bx}{n}
  \SetMathAlphabet\mathbfss{bold}{OT1}{cmss}{bx}{n}
  \ifAMStwofonts
    \ifCUPmtlplainloaded \else
      \DeclareSymbolFont{UPM}{U}{eur}{m}{n}
      \SetSymbolFont{UPM}{bold}{U}{eur}{b}{n}
      \DeclareSymbolFont{AMSa}{U}{msa}{m}{n}
      \DeclareMathSymbol{\upi}{0}{UPM}{"19}
      \DeclareMathSymbol{\umu}{0}{UPM}{"16}
      \DeclareMathSymbol{\upartial}{0}{UPM}{"40}
      \DeclareMathSymbol{\leqslant}{3}{AMSa}{"36}
      \DeclareMathSymbol{\geqslant}{3}{AMSa}{"3E}
    \fi
  \fi
\fi 

\ifCUPmtlplainloaded \else
  \ifAMStwofonts \else 
    \def\upi{\pi}
    \def\umu{\mu}
    \def\upartial{\partial}
  \fi
\fi

\title{Pair Creation by Very High Energy Photons in Gamma-Ray Bursts:
A Unified Picture for Various Energetics of GRBs}
\author[Tomonori Totani]
       {Tomonori Totani \\
        National Astronomical Observatory, Mitaka, Tokyo 181-8588 Japan
        \\
	totani@th.nao.ac.jp}
\date{}

\pubyear{1999}

\begin{document}

\maketitle


\begin{abstract}
The extreme energetics of the gamma-ray burst (GRB)
990123 has revealed that some of GRBs
emit quite a large amount of energy, and total energy release from
GRBs seems to change from burst to burst by 
a factor of $10^2$--$10^3$ as $E_{\rm \gamma, iso} \sim
10^{52-55}$ erg, where $E_{\gamma, \rm iso}$ is the observed
GRB energy when radiation
is isotropic. If all GRBs are triggered by similar events,
such a wide dispersion in energy 
release seems odd. Here we propose a unified picture for these
various energetics of GRBs, in which all GRB events release roughly 
the same amount of
energy of $E_{\rm iso} \sim 10^{55-56}$ erg as relativistic motion,
with the baryon load problem almost resolved. 
A mild dispersion in the initial Lorentz factor ($\Gamma$) results in 
difference of $E_{\rm \gamma, iso}$
up to a factor of $m_p/m_e \sim 10^3$. Protons work as ``a hidden
energy reservoir'' of the total GRB energy, and 
$E_{\rm \gamma, iso}$ depends on the energy transfer
efficiency from protons into electrons (or positrons) in the internal shock. 
We show that this transfer
occurs via $e^\pm$ pair-creation by very high energy photons of
proton-synchrotron radiation, and this efficiency depends quite
sensitively on $\Gamma$. We also show that, in spite of the wide
dispersion in $E_{\rm \gamma, iso}$, this model predicts roughly a constant
photon energy range of the $e^\pm$-pair
synchrotron at $\sim$ MeV, which is well consistent
with GRB observations. 
The optical flash of GRB 990123 can be explained by the internal
shock origin in our model.
The apparent no-correlation between $E_{\rm \gamma, iso}$ and observed
afterglow luminosity is consistent with the expectation of our scenario.

\end{abstract}



\section{Introduction}
Gamma-ray bursts (GRBs) are known as an explosive phenomenon
occurring at cosmological distances ($z \gtilde 1$), emitting an
extremely large amount of energy ($E_{\rm \gamma, iso} \gtilde
10^{51-52}$ erg, when radiation is isotropic)
mostly in the soft $\gamma$-ray range of keV--MeV
(see, e.g., Piran 1999 for a review). 
However, recent observations revealed that
some of GRBs are emitting an even larger amount of energy with 
$E_{\rm \gamma, iso} \gg 10^{52}$ erg. For GRBs 971214 and 980703, 
their total energies were estimated from redshifts of identified 
host galaxies as $E_{\rm \gamma, iso} \sim 3 \times 10^{53}$
and $2 \times 10^{53}$ erg, respectively
(Kulkarni et al. 1998; Djorgovski et al. 1998).
Recently GRB 990123 further pushed up the total energy to $E_{\rm \gamma, iso}
\gtilde 3 \times 10^{54}$ erg ($=1.7 M_\odot c^2$), whose redshift
was estimated as $z = 1.60$ from absorption lines 
observed in its optical afterglow (Kulkarni et al.
1999). It has also been suggested that GRB 980329 might have emitted 
$E_{\rm\gamma, iso} \gtilde
10^{54}$ erg, if the redshift of this GRB is $z \sim 5$ as
inferred from the possible Lyman break observed in the spectrum 
of its optical afterglow (Fruchter 1999). 
On the other hand, there are less energetic bursts,
such as GRB 970508 whose total energy is $E_{\rm \gamma,iso} \sim 
7 \times 10^{51}$ erg (Metzger et al. 1997). Hence, it is now confirmed
that total energy emitted from GRBs is broadly distributed
with a wide dispersion by a factor of $\sim 10^2$--$10^3$,
in spite of the roughly constant photon energy range of $\gamma$-rays.

Currently the most popular explanation for the GRB phenomenon
is dissipation of the kinetic energy of ultra-relativistic bulk motion
with a Lorentz factor of $\Gamma \gtilde 10^{2-3}$, in internal shocks
which are generated by relative velocity difference of relativistic
shells ejected from a central engine (e.g., Piran 1999; see also 
Fenimore et al. 1999b and Sari \& Piran 1999 for recent 
arguments preferring the internal shock scenario to the alternative
external shock).
All the total energy ejected as relativistic bulk motion cannot be
dissipated in internal shocks, and hence the total energy truly emitted as
kinetic motion ($E_{\rm iso}$)
should be larger than the observed $E_{\rm \gamma,iso}$, at
least by a factor of several. Therefore, 
some of GRBs must emit quite a large amount of energy, $E_{\rm iso} \gtilde
10^{55}$ erg. 

There are accumulating evidences for GRBs occurring mostly in star
forming regions of distant galaxies (e.g., Paczy\'nski 1998;
Kulkarni et al. 1998; Bloom et al. 1999),
and it suggests that GRBs are
related to cataclysmic death of massive stars, such as gravitational
collapses or mergers of compact objects. However, if this is the case,
it seems somewhat odd that the total energy emitted from GRBs shows
such a wide dispersion. It may rather be natural
to consider that all GRBs emit roughly the same amount of energy
with $E_{\rm iso} \sim 10^{55-56}$ erg, 
and another physical process is
responsible for the dispersion in the $\gamma$-ray efficiency
($E_{\rm \gamma, iso}/ E_{\rm iso}$),
producing a dispersion in $E_{\gamma, \rm iso}$ by a factor of $\sim 10^3$.
Because of the extremely large amount of energy emitted by GRB 990123,
a strong beaming of $\gamma$-ray emission is now discussed intensively
(e.g., Kulkarni et al. 1999). It is possible that the energy release
from GRB events is roughly the same for all bursts but the beaming factor
significantly changes from burst to burst, resulting in the wide
dispersion in $E_{\rm \gamma, iso}$. However, there is
no plausible reason why the beaming factor changes by a factor of
$\sim 10^3$ from burst to burst, and other explanations may be favorable
in which a relatively small dispersion of a physical parameter of GRBs
results in a factor of $\sim 10^3$ dispersion in $E_{\rm \gamma, iso}$.
In this letter we propose such a model, in which the factor of $\sim 10^3$
is understood roughly as the proton-electron mass ratio and difference
in $\Gamma$ by a factor of $\sim$ 3--4 is the origin of the 
dispersion in $E_{\rm \gamma, iso}$. All GRBs emit the total energy
of $E_{\rm iso} \sim 10^{55-56}$ erg with roughly the same
beaming factor. (Therefore, if GRBs are generated from objects with
stellar masses, a strong beaming with a factor of $b \equiv (4 \pi / \Delta
\Omega) \gtilde 10^{2}$ is highly likely.)

Since the origin of the GRB energy is relativistic bulk motion,
protons should carry a much larger amount of energy than electrons
by a factor of $m_p/m_e \sim 2,000$, 
at least in the initial stage of the internal shock generation.
It is uncertain what fraction of the proton energy is converted into
electrons, but the simplest Coulomb interaction cannot 
transfer the proton energy into electrons within the time scale of GRBs,
as will be shown in this letter. The soft $\gamma$-rays 
are generally considered
to be generated by electrons, because of the short time variability of GRBs.
Therefore it is not unreasonable
that, in some GRBs, only 1/2000 of $E_{\rm iso}$ 
is carried by electrons and then emitted as soft $\gamma$-rays. On the 
other hand, it may also be possible that a physical process works as
an energy conveyor from the hidden energy reservoir (i.e., protons) into
electrons (or positrons). If the energy transfer is almost complete for a GRB, 
a significant fraction of $E_{\rm iso}$ can be radiated as soft $\gamma$-rays.
In our scenario, the energy conveyor is $e^\pm$ pair-creation in internal
shock by very high energy photons radiated by proton-synchrotron.
We will show that the energy transfer efficiency by this process is
quite sensitively dependent 
on $\Gamma$, and hence a relatively small difference in $\Gamma$
from burst to burst gives a large dispersion in $E_{\rm \gamma, iso}$.
We will also show
that the observed photon energy range of the $e^\pm$-pair 
synchrotron is roughly constant at $\sim$ MeV range,
in spite of the large dispersion in $E_{\rm \gamma, iso}$.
This gives an explanation for the universal photon energy range of
GRBs. Otherwise, the photon energy range of $\sim$ MeV 
is difficult to explain by the internal shock scenario with
equipartition magnetic fields,
in which the typical Lorentz factor of particles is $\Gamma^{-1}$ times
smaller than that in external forward shocks.

\section{Pair-Creation in Internal Shocks}
\label{section:pair-cre}
In the following we consider the energy transfer from protons 
into electrons (or $e^\pm$ pairs) 
to be observed as soft $\gamma$-rays, by using a simple
model. The basic model parameters are the following three: 
$E_{\rm iso}$, $\Gamma$, and the duration of GRBs ($T$).
We use the natural units with $c = \hbar = 1$ throughtout this letter.
If the energy emission from the central engine is continuous, 
the duration of energy emission is the same with the observed
GRB duration. If the Lorentz factor of emitted matter has a dispersion
of $\Delta \Gamma \sim \Gamma$, faster ejecta will catch up with the
slower ones and hence internal shocks are generated at a radius
$R \sim \Gamma^2 T$. The energy density at the shock rest 
is given as $\rho_{\rm rest} \sim \rho_{\rm lab}\Gamma^{-2} \sim
E_{\rm iso}/(4 \pi R^2 T \Gamma^2)$,
where $\rho_{\rm lab}$ is the energy density measured in the laboratory
frame. The shock Lorentz factor is of order unity in the internal shock,
and hence the typical Lorentz factor of protons at the shock rest
is also $\gamma_p \sim 1$.
Hence, the particle number density at the shock frame
is given by $n_p \sim n_e \sim \rho_{\rm rest}/m_p$. 

First we show that the Coulomb interaction cannot transport the 
energy carried by protons into electrons to achieve energy equipartition.
The time scale of energy transfer in relativistic plasma is roughly
given by $\tau_{ep} \sim (n_p \sigma_t)^{-1}$, where 
$\sigma_t = 4 \pi L_e (e^2/m_e\gamma_e)^2$ is the transport cross section
for electron-proton collisions and $\gamma_e$ is the electron Lorentz
factor at the shock rest. The Coulomb logarithm
is given by $L_e = \ln (a m_e \gamma_e)$, where $a = 
(m_e \gamma_e/4\pi n_e e^2)^{1/2}$ 
is the Debye length (e.g., Lifshitz \& Pitaevskii 1981). For 
typical parameters, $L_e \sim 40$.
This time scale should be compared with
the expansion time measured in the shell frame, $t_{\rm exp} = R/\Gamma$,
and we find $\tau_{ep}/t_{\rm exp} = 9.3 \times 10^{6} (\gamma_e/10^3)^2
E_{56}^{-1} \Gamma_{300}^5 T_{30}^2 L_{40}^{-1}$, where 
$\Gamma_{300} = \Gamma/300$, $E_{56} = E_{\rm iso}/(10^{56}\rm erg)$,
$T_{30} = T/(30 \rm sec)$, and $L_{40} = L_e/40$. In order for the energy
equipartition between protons and electrons, $\gamma_e$
should become $\sim m_p/m_e \sim 10^3$, but this equation shows that
the time scale of the Coulomb interaction is too long to achieve the
equipartition. Hence it is reasonable to suppose that protons carry
a much larger amount of energy than electrons by a factor of $\sim 10^3$.

Next we consider the synchrotron emission of protons. Proton synchrotron
is generally quite inefficient due to its quite long time scale, but
Totani (1998a, b) has shown that, when $E_{\rm iso}$ is as large as 
$\sim 10^{55-56}$ erg, proton cooling time becomes comparable with
the GRB duration at the acceleration maximum ($\sim 10^{20}$ eV). 
Therefore if the energy spectrum of protons is harder than the
typical shock acceleration spectrum, $dN_p/d\gamma_p \propto \gamma_p^{-2}$,
a considerable fraction of the total energy
can be radiated in the TeV range by synchrotron radiation of protons
accelerated to $\sim 10^{20}$ eV. 
We assume that there is a magnetic field in the internal shock
roughly in equipartition with the turbulent energy density,
i.e., $B^2/(8\pi) = \xi_B \rho_{\rm rest}$, where $\xi_B$ is a parameter
of order unity for the degree of equipartition.
It is generally believed that the time scale of shock acceleration is given
by $t_{\rm acc} =
2 \pi \eta r_L$, where $r_L = m_p \gamma_p / (eB)$ is the Larmor radius and
$\eta$ is a parameter of order unity (e.g., de Jager et al. 1996).
For the condition that 
the maximum proton energy is constrained by the synchrotron cooling,
a relation, $t_{\rm acc} = t_{\rm cool} < t_{\rm exp}$, must hold at
the maximum proton energy,
where $t_{\rm cool} = 6 \pi m_p^3 / (\sigma_T m_e^2 B^2 \gamma_p)$
is the proton-synchrotron cooling time at the shock frame.
The maximum of $\gamma_p$ is determined by $t_{\rm acc}
= t_{\rm cool}$, as $\gamma_{p, \max} = (3 e m_p^2 / \sigma_T m_e^2 B 
\eta)^{1/2}$, and with this $\gamma_{p, \max}$ we can check that
the condition $t_{\rm cool} \ltilde t_{\rm exp}$ actually holds:
\begin{equation}
\frac{t_{\rm cool}}{t_{\rm exp}} = 0.41 \eta^{1/2} \xi_B^{-3/4} E_{56}^{-3/4}
\Gamma_{300}^{7/2} T_{30}^{5/4} \ ,
\label{eq:cooling}
\end{equation}
with reasonable GRB parameters.

Hence the energy carried by protons with $\gamma_p \sim \gamma_{p, \max}$
can efficiently be radiated within the GRB duration.
The synchrotron photon energy
at the shock rest is given by $\varepsilon_{\rm rest} =
\gamma_p^2eB/m_p$, and then one can show that the maximum synchrotron
photon energy corresponding to $\gamma_{p, \max}$
does not depend on $B$ and is expressed only by
fundamental constants, i.e., $\varepsilon_{\max, \rm rest} \sim 3e^2 m_p / 
(\sigma_T m_e^2 \eta) = 46 \eta^{-1}$ GeV. In the following we suggest that
this universal value is responsible for the universality of the
observed photon energy range of GRBs. 
The maximum synchrotron photon energy for observers is then
$\varepsilon_{\rm max, obs} \sim 14 \eta^{-1} \Gamma_{300}$ TeV, 
and this radiation gives an explanation (Totani 1998b) 
for the possible detections of
GRBs above 10 TeV reported by the Tibet (Amenomori et al. 1996) and the HEGRA
(Padilla et al. 1998) groups.

For a typical spectrum of accelerated particles, the luminosity ($\nu
F_\nu$) of synchrotron radiation becomes maximum at the maximum photon
energy. Therefore, energy carried by protons is most efficiently
radiated at around $50\eta^{-1}$ GeV at the shock rest.
However, it must be checked whether these high energy
photons can escape freely from internal shocks. Photons with
$\sim 50 \eta^{-1}$ GeV interact most efficiently with low energy photons at
$2 m_e^2 / (50 \eta^{-1} \rm GeV) \sim 10 \eta$ eV, via the
$\gamma\gamma \rightarrow e^\pm$ reaction whose cross section is
roughly given by the Thomson cross section $\sigma_T$.  Hence radiation
observed as $\sim 3 \Gamma_{300}\eta$ keV 
photons produced by electron synchrotron 
radiation in the internal shock could be a significant
absorber of the very high energy photons. 
Electrons carry about 
$(m_e/m_p$) of the total energy, and suppose that a fraction
$\zeta$ of the total electron energy
is converted into target photons with restframe energy around
$\varepsilon_{\rm t, rest} \sim 10 \eta$ eV. 
The column density of the low energy target photons then becomes 
\begin{equation}
N_\gamma \sim \frac{\zeta E_{\rm iso} m_e}{4 \pi R^2 m_p \Gamma
\varepsilon_{\rm t, rest}} \ .
\end{equation}
Then the opacity of very high energy photons is given as
$\tau_{\gamma \gamma} \sim \sigma_T N_\gamma \sim 0.9 \zeta_{-2} E_{56}
T_{30}^{-2} \Gamma_{300}^{-5} \eta^{-1}$, where $\zeta_{-2} =
\zeta/10^{-2}$. We will discuss the value of $\zeta$ and electron
synchrotron radiation in more detail in \S \ref{section:electron}.

It should be noted that $\tau_{\gamma \gamma}$
is of order unity, and also quite sensitively
depends on $\Gamma$. When $\Gamma \sim 300$, most of 
proton-synchrotron photons create $e^\pm$ pairs due to the optical
depth of order unity. Therefore energy conversion from protons into 
$e^\pm$ pairs is efficient, and synchrotron radiation of the created
pairs produces the energetic GRBs such as GRB 990123 or 980329.
On the other hand, with increasing $\Gamma$ above $\sim 300$, the energy
transfer into $e^\pm$ pairs becomes rapidly inefficient, because
$\tau_{\gamma \gamma}$ decreases as $\propto \Gamma^{-5}$, and
also because the 
proton acceleration becomes limited by the expansion time,
rather than the synchrotron cooling (see eq. \ref{eq:cooling}).
In this case $E_{\rm \gamma, iso}$ is expected to be much smaller than
$E_{\rm iso}$, down to $\sim (m_e/m_p)E_{\rm iso}$ which is the 
energy released by the original electron synchrotron.
A modest variation in $\Gamma$ ($\sim$ a 
factor of 3--4), which is necessary for the internal shock scenario of
GRBs, will result in difference of $\tau_{\gamma \gamma}$ by a factor
of $\sim 10^3$,
and hence explains the observed dispersion in $E_{\rm \gamma, iso}$.

Then our next interest is whether the created pairs produce the 
soft $\gamma$-rays as observed, in the correct photon energy range.
From energy conservation, it is clear that each of created pairs has
roughly a half of the energy of very high energy photons ($\sim 50
\eta^{-1}$ GeV), and hence
the Lorentz factor of pairs in the shock rest is typically $\gamma_\pm 
\sim 50 \eta^{-1} {\rm GeV} / (2 m_e) \sim 4.9 \times 10^4 \eta^{-1}$.
The observed pair-synchrotron photon energy is given by $\varepsilon_{
\rm obs} = \Gamma \gamma_\pm^2 eB / m_e = 5.1 \eta^{-2} \Gamma_{300}^{-2}
\xi_B^{1/2} E_{56}^{1/2} T_{30}^{-3/2} \ \rm MeV$. Considering the
simplicity of the model, this photon energy
is well consistent with the observed $\nu F_\nu$ peaks of GRBs
at around 0.1--1 MeV (e.g., Mallozzi et al. 1995), 
and hence this peak can be interpreted as the synchrotron photon
energy corresponding to the typical energy of created pairs.
The $\nu F_\nu$ peak depends relatively weakly on $\Gamma$,
because of the universal value of the maximum proton-synchrotron
photon energy.
A dispersion by a factor of $\sim$ 3--4 in $\Gamma$ would produce change
in the $\nu F_\nu$ peak at most a factor of 10, which is consistent
with the observation (Mallozzi et al. 1995). Therefore, we have 
successfully explained the fact that $E_{\rm  \gamma, iso}$
changes almost by a factor of $10^3$ from burst to burst, 
while the photon energy range of GRBs shows little change.

\section{Electron Synchrotron and Optical Flash}
\label{section:electron}
Here we consider the synchrotron radiation of electrons
which are originally loaded in the ejected matter (i.e., not 
created pairs). These electrons must produce the target photons
which interact with the very high energy proton-synchrotron photons.
The shock Lorentz factor in internal shocks is of order unity, and
hence the minimum Lorentz factor of electrons at the shock rest is
$\gamma_e \sim 1$, because there is almost no energy transfer from protons
into electrons in our scenario. 
In this case, the observed electron-synchrotron radiation
starts at a very low photon energy of 
\begin{equation}
\varepsilon_{\gamma, \rm obs} = 
\Gamma \gamma_e^2 e B / m_e \sim 2.1 \times 10^{-3} \gamma_e^2 
\Gamma_{300}^{-2} \xi_B^{1/2} E_{56}^{1/2} T_{30}^{-3/2} \ \rm eV \ .
\label{eq:e-sync}
\end{equation}
The synchrotron cooling time of electrons for observers is 
$t_{\rm obs} = t_{\rm rest}/(2 \Gamma)$, where $t_{\rm rest}
= 6 \pi m_e / (\sigma_T B^2 \gamma_e)$ is the cooling time at the shock frame. 
Comparing the cooling time for observers with
the GRB duration ($T$), we find that the electron synchrotron is in
the efficient cooling regime above an observed photon energy of
$\varepsilon_{\rm cl} = 2.8 \times 10^{-5} \xi_B^{-3/2} E_{56}^{-3/2}
\Gamma_{300}^8 T_{30}^{5/2}$ eV. 
The synchrotron self-absorption should also be taken into
account. By using the previously estimated electron number density
and magnetic fields in the internal shock, the optical depth ($\tau_{\rm 
syn} \sim \chi T \Gamma$)
of the synchrotron self-absorption can be estimated (e.g., Longair 1994),
where $\chi$ is the absorption coefficient of the self-absorption and
$T \Gamma$ is the shell thickness measured in the shock frame.
From this optical depth, we find that the synchrotron radiation
becomes optically thin above an observed photon energy
of $\varepsilon_{\rm ab} = 5.6 \tau_{\rm syn}^{-1/3} \xi_B^{1/3}
\gamma_{e, \min}^{1/3} E_{56}^{2/3}
T_{30}^{-5/3} \Gamma_{300}^{-8/3}$ eV, where we have assumed that
the spectrum of electrons as $dN/d\gamma_e \propto \gamma_e^{-2}$ and
$\gamma_{e, \min}$ is the minimum value of $\gamma_e$.
Therefore efficient electron-synchrotron emission starts at
the optical band, where emission is marginally optically thin
depending the model parameters.
The energy of 
target photons for the very high energy proton-synchrotron photons
is $\sim 3 \Gamma_{300} \eta$ keV for observers, 
and in this band the radiation is optically thin and 
in the efficient cooling regime. Therefore most
of energy carried by electrons corresponding to the target photon energy
is converted into the target photons. The energy 
fraction of such electrons in the
total electron energy depends on the acceleration spectrum, 
and $\zeta \sim 10^{-2}$--$10^{-1}$ might be a plausible value. 

It should be noted that 
the electron-synchrotron radiation
extends to lower photon energies of the optical range. The optical
flash observed for GRB 990123 (Akerlof et al. 1999)
can be attributed to this internal shock
origin, although this does not reject the explanation by reverse shocks
(Sari \& Piran 1999; M\'esz\'aros \& Rees 1999). 
The peak optical flux of $5 \times 10^{49} \ \rm erg \ s^{-1}$
and the flash duration of $\sim 50$ sec suggest that the energy emitted as
the optical flash is about $E_{\rm opt, iso}
\sim 10^{51}$ erg, which is about
$10^3$ times smaller than the energy emitted as $\gamma$-rays. 
In our scenario, this difference can be understood roughly as the
proton-electron mass ratio,
because the optical flash is synchrotron radiation of electrons originally
loaded in the internal shock matter, while $\gamma$-rays are synchrotron
radiation of created $e^\pm$ pairs which possess a significant fraction of 
energy originally carried by protons. 

This scenario predicts that the total energy emitted as optical
flashes is roughly constant at $E_{\rm opt, iso} \sim 10^{51-52}$ erg,
while $E_{\rm \gamma, iso}$ considerably
changes from burst to burst due to different efficiencies of $e^\pm$ creation.
This trend is different from that in other GRB models which predict
a rough scaling of $\gamma$-ray/optical fluence, and hence future
observations may discriminate these different scenarios.
In fact, the observed correlation between $E_{\gamma, \rm iso}$ and
$\gamma$-ray/X-ray peak flux ratio (Norris, Marani, \& Bonnell
1999) may already suggest this trend. Our scenario predicts
that variation in X-ray luminosity is also small compared with that 
in $\gamma$-ray
luminosity, because the synchrotron radiation of created pairs
is radiated mainly in the $\gamma$-ray range, as we have shown. 
Therefore $\gamma$/X ratio should increase with $E_{\gamma, \rm iso}$.
For the five GRBs with redshift information (including GRB 980329),
$E_{\rm \gamma, iso}$ = 0.55, 11, 26, 110, 330 $\times 10^{52}$ erg
for GRBs 970508, 980703, 971214, 980329, and 990123, respectively.
[Here and in the following 
we assume the cosmological parameters as $(h, \Omega_0,
\Omega_\Lambda) = (0.7, 0.3, 0.7)$.]
The $\gamma$/X peak flux ratio for these GRBs is 25, 40, 56, 120, and
252 in the same order (Norris et al. 1999). Although scatter is
considerable, such a trend is clearly seen. The $\gamma$/X ratio 
does not increase in proportion to $E_{\rm \gamma, iso}$, and this
can be understood by the contamination of pair-synchrotron radiation
in the X-ray band. The cooling time of pairs is much shorter than 
the GRB duration, and synchrotron photon energy should decrease with
the cooling of pairs, making some contribution in the X-ray band.

Similarly our model predicts that the energy emission from afterglows
is not correlated with $E_{\gamma, \rm iso}$, because 
roughly the same amount of energy $\sim E_{\rm iso}$ is injected
into interstellar matter for all GRBs regardless of the 
pair-creation efficiency. 
We note that such a trend has already been seen in the
flux ratio of GRBs and optical afterglows. We can estimate 
a characteristic amount of energy emitted from afterglows, 
as $E_R(t) \equiv
\nu_R L_\nu(\nu_R) t$, where $\nu_R$ is the R-band frequency,
$L_\nu$ the afterglow luminosity per unit frequency,
$t$ the time from the burst, and all these quantities
are those measured in the rest frame of a host galaxy. 
By using the observed R-band light curves (Pian et al. 1998;
Kulkarni et al. 1998; Galama et al. 1999), we have estimated
$E_R$ for the famous three GRBs with known redshifts:
$E_R({\rm 1day}) = 
4.2, 0.9,$ and $1.2 \times 10^{49}$ ergs
for GRB 970508, 971214, and 990123, respectively. Here we have assumed
the spectrum and time evolution of afterglows as $f_\nu \propto t^{-\alpha}
\nu^{-\beta}$, with $(\alpha, \beta)$ = (1.1, 0.7) which are typical
values for GRB afterglows, to take into account the K-correction and
cosmological time dilation. Clearly there is almost no correlation between
$E_{\gamma, \rm iso}$ and afterglow luminosity, and the
latter seems rather uniform. Such a trend 
is consistent with the expectation of our model.

\section{Structure of the Emission Region in Internal Shocks}
In this section we consider the size of emission region where created 
$e^\pm$ pairs
are injected, and we show that such emitting regions are likely to
be clumpy. Once pair creation by very high energy photons starts
at a seed region, created pairs radiate photons mainly in the observer's
soft $\gamma$-ray band, but also radiate lower-energy 
target photons at $\sim$ keV when the Lorentz factor of pairs decreases
due to cooling. Increase of target photons will be 
followed by more efficient pair creation, and then be followed again by
increase of target photons. Therefore perturbation in local
pair-creation rate will unstably glow to a clumpy structure of $\gamma$-ray
emitting region. The created pairs
cannot freely expand by the speed of light, but they are trapped around
the seed region due to magnetic fields. This picture is very similar
to ``the seed growth scenario'' discussed in Fenimore et al. (1996),
and it may give an explanation for the
very low surface filling factor of GRBs (Fenimore et al. 1996, 1999a).
If emission region is homogeneous with a scale of visible shock region,
complicated time structure seen in GRB light curves cannot be explained.
Rather, only very small fraction (i.e., the surface filling factor)
of shocked region can be active for $\gamma$-ray emission. 
It should be noted that, in the internal shock scenario, 
the stochastic nature of internal shock generation may also provide
an origin of the inhomogeneity in the shocked region, resulting in
the small filling factor. However, the scale and amplitude of 
the inhomogeneity of internal shocks are highly uncertain, and our
model provides another candidate for the origin of the small filling factor.

It is difficult to theoretically 
estimate a typical size of pair-creation regions,
but the minimum size possible may be estimated by a diffusion process
with a mean free path of order the Larmor radius of pairs.
The Larmor radius 
for the typical pair energy of $\sim 25 \eta^{-1}$ GeV at the shock rest
is $r_L \sim 1.4 \times 10^5 \eta^{-1} \xi_B^{-1/2} E_{56}^{-1/2}
\Gamma_{300}^3 T_{30}^{3/2}$ cm. Then the size achieved by a random diffusion
process within the restframe expansion time ($t_{\rm exp} \sim R/\Gamma$)
becomes $\Delta r_{\rm
em} \sim (r_L R \Gamma^{-1})^{1/2}$ at the shock rest.
The radial size of this emitting region is $\Delta r_{\rm em}/\Gamma$
in the laboratory frame, and this gives a time scale of a pulse duration
as $T_p \sim \Delta r_{\rm em}/\Gamma$ (e.g., Fenimore et al. 1996).
Then we obtain the ratio of a pulse width to the total duration
as $T_p/T = (r_L/T \Gamma)^{1/2} \sim 2.2 \times 10^{-5}
\eta^{-1/2} \xi_B^{-1/4} E_{56}^{-1/4} \Gamma_{300} T_{30}^{1/4}$. 
Although this estimate may be too simple, 
this is sufficiently small to explain the complicated time structure
seen in long-duration GRBs. 

Each pair-creation region makes a strong
pulse in the GRB light curve. If only one of such pulses has a peak
flux above a detection threshold of a detector, the observed duration of
GRBs will be much shorter than the true duration $T$. This may be
responsible for the bimodal distribution of the observed
GRB durations (Kouveliotou et al. 1993). The short duration peak
may correspond to a typical size of a pair-creation region, while the long 
duration peak corresponds to the energy-emission time scale from
the central engine.

We speculate that GRBs with complicated
time structure are those in which the pair-creation process is efficient
(and hence energetic),
while GRBs with smooth structure, such as ``FRED'' GRBs (Fast Rise and
Exponential Decay, see, e.g., Fenimore et al. 1996), are 
those with inefficient pair creation process (and hence less
energetic). Norris et al. (1999)
reported that there is a correlation between $E_{\rm \gamma, iso}$
and time lags in energy-dependent light curves of GRBs. 
Most GRBs show short time lags ($\ltilde 0.1$ sec), 
and energetic GRBs such as GRB 980329 or 990123 fall
in this category.
On the other hand, a fraction of GRBs including 
less energetic bursts such as
GRB 970508 show longer time lags (Norris et al. 1999, see also
Band 1997).
The energy dependent time lags can be understood by delayed arrivals
of lower energy photons from off-axis regions compared with the hard
onset of a pulse on the line-of-sight to the source (Norris et al. 1999).
Therefore, the time lags are expected to increase with the surface
filling factor. In our scenario,
energetic GRBs in which the pair creation is the dominant process
of $\gamma$-ray emission
should show smaller time lags, because each emission region
is very compact compared with the size of internal shocks.
In less energetic bursts,
synchrotron radiation by originally 
loaded electrons becomes more significant due to inefficient pair
creation, and relatively homogeneous distribution of original electrons
will emit $\gamma$-rays with longer time scales. We suggest that this
may be the origin of the correlation observed by Norris et al. (1999).

\section{Discussion}
In this letter we have suggested that not only GRB 990123, but also
all GRBs emit a large total energy of $E_{\rm iso} \gtilde 10^{55}$ erg.
We note that the baryon load problem, which has been a quite difficult
problem for a long time in theoretical modeling of GRBs, is considerably
relaxed. The baryon mass which should be loaded in the relativistic
bulk motion is $m_{\rm iso} \sim E_{\rm iso}/\Gamma \sim 0.2 
E_{56}\Gamma_{300}^{-1}M_\odot$ for $4\pi$ steradian.
This value is still small,
but seems not unreasonable in stellar death models. On the other hand,
if we consider the conventional energy scale of $E_{\rm iso} \sim
10^{52}$ erg, immediately we come up against the unreasonably small
baryon load of $m_{\rm iso} \sim 10^{-5} M_\odot$. Therefore our scenario 
makes the theoretical modeling of GRBs considerably easier in a sense
that the baryon load problem is almost resolved, 
although the energy release required
is quite large and a strong beaming is necessary for stellar death models.

Brainerd (1998a, b) argued that the total energy released
by GRB 970508 should be much larger than that observed as 
$\gamma$-ray fluence, if the radio afterglow of this GRB is
optically-thick synchrotron radiation of electrons. This is quite a
model-independent argument based on the source size of the radio 
afterglow ($\sim 3 \mu$as) inferred from the time variability of radio flux
due to the scintillation by the interstellar matter within the Galaxy
(Frail et al. 1997). This analysis may give a support for our scenario
in which the true kinetic energy release of GRB 970508 is the same with
GRB 990123.

The time profile of GRB 990123 suggests that the $\gamma$-ray emission
of GRBs is not the external shock origin (Fenimore et al. 1999b).
The advantage of the external forward 
shock scenario for the $\gamma$-ray emission
was that typical photon energy range of electron synchrotron 
is much higher than that in internal or reverse shocks.
The shock Lorentz factor of the external forward
shock is the same with that of the
relativistic bulk motion ($\Gamma$), 
while it is of order unity for the internal
or reverse shocks. Therefore, the particle Lorentz factor in external forward
shocks is also $\Gamma$ times larger than that in the internal or reverse
shocks, and hence the synchrotron photon energy is $\Gamma^2$ times 
higher than the other two. In fact, as shown in eq. (\ref{eq:e-sync}), typical 
photon energy band of electron-synchrotron from internal shocks
is much lower than the
observed emission band of GRBs ($\sim$ MeV), 
when typical $\gamma_e$ is $\sim$1.
Even when the energy conversion from protons into electrons is efficient,
i.e., $\gamma_e \sim 10^3$, the typical synchrotron photon energy is
at most $\sim 10$--100 eV. (Note that in this case
we should use $E_{\rm iso} \sim E_{\gamma, \rm iso} 
\sim 10^{52-53}$ erg for typical GRBs.)
If we attribute the optical flash to the
reverse or internal shocks (Sari \& Piran 1999; M\'esz\'aros \& Rees 1999), 
and $\gamma$-ray emission is not the external
shock origin, we have a difficulty in explaining the typical photon
energy of the $\gamma$-ray emission of GRBs with equipartition
magnetic fields. 
As we have shown in this letter, the pair creation by very high energy
photons of proton-synchrotron provides a promising candidate for
the explanation of the $\gamma$-ray energy range of GRBs.

The simplest prediction of our scenario is that we should observe
a comparable, or even larger amount of energy emission in the TeV range
compared with $E_{\rm \gamma, iso}$, from some fraction of GRBs
in which the pair-creation optical depth
$\tau_{\gamma\gamma}$ is of order unity or less.
On the other hand, when $\tau_{\gamma \gamma} \gg 1$, the TeV flux should be 
strongly attenuated.
Therefore it is difficult to predict the
flux ratio between TeV and MeV ranges, and it will considerably
change from burst to burst.
But if a larger amount of energy emission in TeV range
than $E_{\rm \gamma, iso}$ is observed for a GRB in the future, 
it would give a strong support for our scenario.
Such strong TeV bursts may already have been
detected by the Tibet (Amenomori et al. 1996) and the HEGRA groups 
(Padilla et al. 1998).


\end{document}